# Separation of sinusoidal and chirp components using Compressive sensing approach

Zoja Vulaj, Faris Kardović
Faculty of Electrical Engineering
University of Montenegro
Podgorica, Montenegro

*Abstract*—**In this paper we deal with the linear frequency modulated signals and radar signals that are affected by disturbance which is the inevitable phenomenon in everyday communications. The considered cases represent the cases when the signals of interest overlap with other signals or with noise. In order to successfully separate these signals we propose the compressive sensing method, which states that the useful signal part can be separated successfully from a small amount of measurements as long as the acquired signal can be presented as sparse in a certain transformation domain. The effectiveness of our approach is proven experimentally through examples.**

## I. INTRODUCTION

The development of technology and thereby the development of a variety of devices has led to the need of performing a fine sampling of the acquired signal. The first approach of reconstructing signals from measured data was based on the Nyquist-Shannon sampling theorem, which states that the sampling rate must be twice the highest frequency. Nowadays, this sampling may become too high in many applications. Therefore, a new method of signal processing, named CS, was developed [1]. Using CS, signals can be successfully reconstructed from a small amount of measurements, which decreases the storage demand and makes the transmission of the signal easier [2], [3]. Since its development, CS has found a wide range of application, including communication networks and radar signal processing [1]. Signals in real application, are often disturbed by unwanted electromagnetic interference from other transmitters, deliberate disturbances or other phenomena that affects the useful information and disrupts the information flow. If the signal and the disturbance components highly overlap in time and frequency, the separation of these components using the conventional methods of filtering or windowing will not give desirable results. Furthermore, if these components reside over common time-frequency regions or in the cases when the disturbance is much stronger than the desired signal in the overlapping regions, the separation becomes more complex. The useful signal part and the undesired disturbance can be separated using CS [4]. The signal can be reconstructed using CS as long as it is sparse or can become sparse after a certain transformation. A signal is sparse when it is represented as a linear combination of a few components in a particular basis. Different signals can be presented as sparse in different transformation domains. Consider the case of the air surveillance radar whose task is to scan several kilometers in range and detect targets. Most of the time, a few or no targets will be present over time. This means that the target response can be considered sparse in the time domain. Most signals, including the signals of our interest, can be presented as sparse in the time-frequency domain [5], [6]. A typical example is chirp signal used in radar with changing frequencies. The time-frequency analysis in these cases can successfully handle the nonlinearity and nonstationarity of the data. In this paper, we assume that the signals in communications and radar signals, which are of the form of sinusoids, are highly disturbed by other signals or by noise. As a consequence, the available samples represent the contribution of both the useful signal part and the disturbance, which leads to the need of introducing the L-statistics that applies sorting and discarding of samples which are regarded to belong to the noise. The missing samples, using the remaining ones, are reconstructed using CS. As a result we get the useful sinusoidal signal. In both communications and radar the disturbance is often a chirp waveform signal. In radar, except chirp signals, the disturbance might be also the sinusoidally modulated signals [7]. Performing the same procedure, both types of disturbances can be eliminated and the useful signal part can be successfully reconstructed. The presented approach of signal separation is also used for the separation of signals of two different wireless standards (Bluetooth and IEEE 802.11b standard) [8]. To get a better picture of the application of CS separation in everyday communication we imagine two radars, one sending sinusoids, while the other sends chirp signals. Both signals are received as overlapping and as such, they have to be separated. CS can be applied also to separate multicomponent signals received by only one receiver which treats them as one single signal. Another application includes systems that consist of multiple antennas which transmit signals to a targeted location where the reflected echoes are received and processed [9]. Chirp signals have been extensively used in radar and sonar systems to determine distance, speed and object detection [6]. In wireless communication, this approach can be used to monitor the network itself, to adjust the data transmitting behavior by ensuring that it does not cause interference to other primary users of the frequency. It can also be used in data separation when the base station receives the summation of the randomly projected data transmitted simultaneously in a phase-synchronized channel by different sensors.

The paper is organized as follows. The theoretical background on CS is given in section 2. In section 3 are introduced two cases of signal separation (the first one based on STFT and the second based on the LPFT) and the reconstruction algorithms. The experimental results are analyzed in section 4. In section 5 are given the concluding remarks.

## II. THEORETICAL BACKGROUND ON COMPRESSIVE SENSING FOR SIGNAL SEPARATION

CS is used to reconstruct signals that are sparse in a certain transformation domain. According to this, the signal of interest $x$ is not sampled in $N$ samples relying on the Nyquist-Shannon theorem, but is a signal presented with $M$ samples that are the acquired measurements ($M < N$) [10]-[13]. These measurements can be obtained using a CS matrix $\Omega$. The corresponding incomplete set of samples can be presented as a measurement vector which can be expressed as:

$$y = \Omega x. \quad (1)$$

The elements of the matrix $\Omega$ make possible the reconstruction of a full data set of $N$ samples out of $M$ available samples. Knowing that the acquired signals sparsity is the main requirement of CS, it can be approximated by a small number of non-zero coefficients in a suitable transformation domain. Consequently, the signal $x$ can be expressed:

$$x = T\lambda, \quad (2)$$

where T is the matrix of size $N \times N$ which represents the signal in a certain transformation basis and $\lambda$ is the vector which contains the transform coefficients. If the number of non-zero coefficients in $\lambda$ is $C < N$, we may say that the signal $x$ is $C$ sparse in the domain T. Now, the signal $y$ can be obtained as:

$$y = T\Omega\lambda. \quad (3)$$

## III. CS BASED COMPONENTS SEPARATION

In communications a narrowband signal can be disturbed by a frequency hopping jammer of shorter duration than the considered time interval or it might overlap with narrowband signals within the same interval.

### A. Case 1

Consider the case of a stationary sinusoidal signal corrupted by a non-stationary chirp signal:

$$y(n) = x(n) + e(n). \quad (4)$$

The Discrete Fourier Transformation (DFT) of the signal $y(n)$ is defined as:

$$Y(k) = X(k) + E(k), \quad (5)$$

where $X(k) \neq 0$ for $k \in \{k_1, k_2, ..., k_K\}, K < N$ ($N$ is the number of time samples). Signals in communication are usually time-varying, which means that the signals frequency varies over time. This implies that we will get more information if we picture the signal in time-frequency representation. In this case we propose the simplest time-frequency representation, which is the short-time Fourier transform (STFT) [7]. The STFT is calculated using a rectangular window of the width $M$:

$$\Gamma(n,k) = \sum_{m=0}^{M-1} y(n+m)e^{-j2\pi mk/M}. \quad (6)$$

After the values of the STFT are sorted for each frequency $k$ along the time axis $n$, if we remove the strongest values for each frequency, the entire or most of the disturbance components are omitted. If we assume that after this step $M$ values are left, by summing these values we will get the spectrum of the desired signal as a result. For a given frequency $k$, one STFT column with $M$ elements is:

$$S_k(m) = sort\{\Gamma(m,k), m = 0, ..., M-1\}, \quad (7)$$

where $|S_k(1)| \leq ... \leq |S_k(M-1)|$. If we use the commutative property of addition, then:

$$\sum_{m=0}^{M-1} \Gamma(m,k) = \sum_{m=0}^{M-1} S_k(m) = S(k). \quad (8)$$

If we omit $Q$ highest values and $P$ smallest values of $S_k$ for each $k$, based on L-statistics, as a result we get an estimate of $S(k)$ expressed as follows:

$$S_L(k) = \sum_{m=P}^{M-Q} S_k(m). \quad (9)$$

The L-statistics implies the elimination of a part of measured data, before further analysis. Based on its rules, when the signal is corrupted by a disturbance at a certain frequency in its STFT representation, the highest values along that frequency belong to the interference. The highest values in the STFT also represent the overlapping points in the case when, at a certain frequency, there is contribution from both the desired components and the interference. Some of the lowest values, along with the highest, need to be removed in order to eliminate the unwanted disturbance. If we subtract the reconstructed signal from the initial signal, as a result we will get the chirp signal.

### B. Case 2

Assuming that, in this case, the signal $y(n)$ is a result of not only the contribution of the stationary sinusoidal signal and the non-stationary chirp signal, but also of the unintentional noise, it can be expressed as follows:

$$y(n) = x_1(n) + x_2(n) + e(n), \quad (10)$$

If we apply the approach proposed in Case 1, after the time-frequency points belonging to the stationary signal are removed from the time-frequency representation of the signal $y(n)$, as a result we will get the time-frequency representation of the chirp signal corrupted by the noise. For data local behavior revealing, when dealing with chirp signals, the Local Polynomial Fourier Transform (LPFT) is proposed [7], [14], rather than the Short Time Fourier Transformation (STFT) which provides only low resolution for time-varying signals. The LPFT represents a generalization of the STFT that uses extra parameters to approximate the instantaneous frequency characteristic, and thus, provides a much better concentration and resolution.

Using the L-statistics, time-frequency regions corresponding to the non-stationary disturbance over all windows are identified and removed from consideration. The sparse signal will be recovered authentically based on the set of the remaining time-frequency points. If the considered chirp signal $x_2$ is sparse in the frequency-chirp rate domain, its LPFT using a rectangular window of the width $M$ in the matrix form can be written as:

$$F_M x_2(n) = \begin{pmatrix} \Pi(n,0) \\ \Pi(n,1) \\ \vdots \\ \Pi(n,M-1) \end{pmatrix} = \begin{pmatrix} F_M & 0 & \cdots & 0 \\ 0 & F_M & 0 & 0 \\ \vdots & \vdots & \ddots & \vdots \\ 0 & 0 & \cdots & F_M \end{pmatrix} \begin{pmatrix} x_2(n) \\ x_2(n+1) \\ \vdots \\ x_2(n+M-1) \end{pmatrix}. \quad (11)$$

where $x_2(n) = f(n)e^{-j\alpha n^2}$. The exponential function is used to demodulate the signal, i.e. to compensate the linear frequency modulated part of the signal ($\alpha$ is the chirp rate). $F_M$ is the $M \times M$ DFT matrix of elements:

$$F(m,k) = e^{-j2\pi km/M}, m=0,\ldots,M-1, k=0,\ldots,M-1. \quad (12)$$

If the windows used for LPFT are non-overlapping, after calculating all LPFT vectors, they can be combined as:

$$\Pi = F_{M,N} x_2, \quad (13)$$

where $F_{M,N} = I_{N/M} \otimes F_M$ is a matrix of the size $N \times N$ obtained as a Kronecker product of the identity matrix of size $(N/M) \times (N/M)$:

$$F_{M,N} = \begin{pmatrix} 1 & 0 & \cdots & 0 \\ 0 & 1 & \cdots & 0 \\ \vdots & \vdots & \ddots & \vdots \\ 0 & 0 & \cdots & 1 \end{pmatrix} \begin{pmatrix} F_M & 0 & \cdots & 0 \\ 0 & F_M & \cdots & 0 \\ \vdots & \vdots & \ddots & \vdots \\ 0 & 0 & \cdots & F_M \end{pmatrix} = \begin{pmatrix} F_M & 0 & \cdots & 0 \\ 0 & F_M & \cdots & 0 \\ \vdots & \vdots & \ddots & \vdots \\ 0 & 0 & \cdots & F_M \end{pmatrix}. \quad (14)$$

The identity matrix is used to ensure that only elements belonging to the diagonal are retained while other elements are set to 0. Equation (16) written in the matrix form can be presented as:

$$\begin{pmatrix} \Pi_M(0) \\ \Pi_M(M) \\ \vdots \\ \Pi_M(N-M) \end{pmatrix} = \begin{pmatrix} 1 & 0 & \cdots & 0 \\ 0 & 1 & \cdots & 0 \\ \vdots & \vdots & \ddots & \vdots \\ 0 & 0 & \cdots & 1 \end{pmatrix} \begin{pmatrix} F_M & 0 & \cdots & 0 \\ 0 & F_M & \cdots & 0 \\ \vdots & \vdots & \ddots & \vdots \\ 0 & 0 & \cdots & F_M \end{pmatrix} \begin{pmatrix} x_2(n) \\ x_2(n+1) \\ \vdots \\ x_2(n+M-1) \end{pmatrix}. \quad (15)$$

If we calculate the DFT of $x_2$ and denote it as $X_2$, the LPFT can now be written as:

$$\Pi = F_{M,N} F_N^{-1} X_2, \quad (16)$$

$F_N^{-1}$ is the DFT matrix of size $N \times N$. As explained in Case 1, the spectrum of the signal after the time-frequency points that represent the disturbance are omitted results in a new spectrum of the signal that is free of disturbance. Follows, when we apply the L-statistics to the LPFT, as a result we will get a new LPFT vector in time, which, after the disturbance components are removed, forms the LPFT whose components are the remaining CS coefficients. From all available LPFT values, for all frequencies, we will form a vector denoted as $\Pi^{CS}$. The corresponding CS matrix is formed by omitting the rows in $F_{M,N} F_N^{-1}$ that correspond to the removed positions in the $\Pi^{CS}$ vector. The decompression of the signal is performed using reconstruction algorithm [11], [15]:

$$\min \|X_2\|_{\ell_1} \text{ subject to } \Pi^{CS} = F^{CS} X_2. \quad (17)$$

As a result we get the reconstructed DFT vector $X_2$ which obtains the time domain signal $x_2(n)$. By performing the opposite of the procedure we used to demodulate the signal, which is to multiply the time domain signal with the factor $e^{-j\alpha n^2}$ we have re-modulated the signal to the original signal.

## IV. EXPERIMENTAL RESULTS

*Example 1:*

Consider a sparse signal corrupted by one component (one chirp signal). The STFT of the signal is calculated and presented in Fig 1a. In Fig 1b is presented the STFT after sorting its values, Fig 1c shows the STFT after applying the L-statistics and removing 50% of the coefficients. The reconstructed STFT of the useful signal and the disturbance are shown in Figs 1d and 1e. The Fourier Transform of the initial signal and the Fourier Transform of the disturbance are shown in Fig 2 (first and third row). In Fig 2 (second row) is presented the useful signal, which is a result of removing the disturbance components form the FFT of the initial signal.

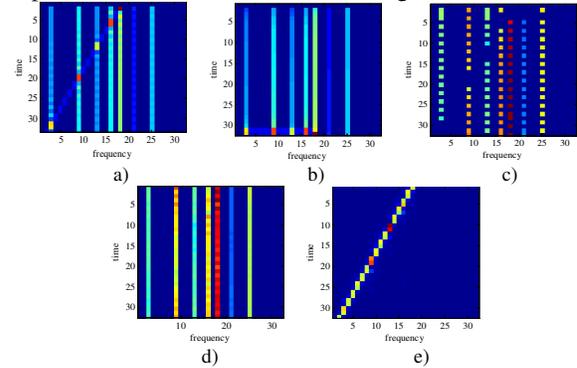

Figure 1. STFT of the original signal, sorted STFT, STFT of the signal after eliminating the coefficients (first row, from left to right), reconstructed STFT of the useful part of the signal, the disturbance (second row, from left to right)

*Example 2:*

Let us now consider a sparse signal corrupted by two disturbances presented as chirp signals. The STFT of the useful signal and its sorted values are shown in Fig 3a and Fig 3b. After applying the L-statistics and removing the disturbance components the STFT looks as presented in Fig 3c. The disturbance components are presented in Fig 3d. In Fig 3e is shown the STFT of the useful part of the signal. The reconstruction is performed using 45% components. The Fourier transform of the initial signal is shown in Fig 4(first row), while in Fig 4(second row) is presented the Fourier transform of the useful signal after the disturbance is removed.

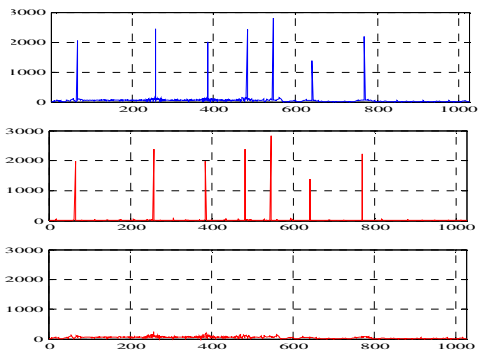

Figure 2.  FFT of the initial signal (first row), FFT of the useful signal (second row), FFT of the disturbance (third row)

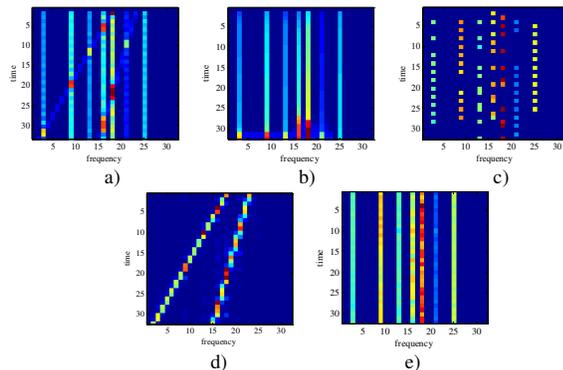

a)          b)          c)

d)          e)

Figure 3.  STFT of the useful signal, sorted STFT, the STFT of the remaining signal after the elimination of coefficients (first row, from left to right), STFT of the disturbance, STFT of the useful part of the signal (second row, from left to right).

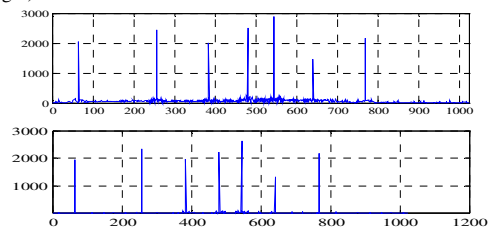

Figure 4.  FFT of the initial signal (first row), FFT of the useful signal (second row).

## V. CONCLUSION

The L-statistics based separation of the signals in the TF domain is considered in the paper. Calculating TF distribution of the original signal and sorting its values, it is possible to separate regions that belong to signal from unwanted components. Full TF of the signal is obtained by using CS reconstruction. We have observed cases when the signal is sinusoidal and unwanted signal is in the chirp form. It is shown that this method can separate useful signal no matter how many components are in the signal and how many chirps are present. Also, theoretical background on the case when sinusoids and chirps are part of the signal and noise is unwanted component is presented. In such cases useful signal parts can be extracted in two steps. Firstly, sinusoids are extracted using the described method. Second step involves chirp demodulation by using LPTF to obtain sinusoids, and then L-statistics is applied on such sinusoidal signal to separate it from noise.

## VI. ACKNOWLEDGEMENT

The authors are thankful to Professors and assistants within the Laboratory for Multimedia Signals and Systems, at the University of Montenegro, for providing the ideas, codes, literature and results developed for the project CS-ICT (funded by the Montenegrin Ministry of Science).